\documentclass[doublecol]{epl2}
\title{Influence of gluon  behavior on heavy-quark pair production }% Force line breaks with \\
\author{G.R.Boroun\inst{1} \and B.Rezaei\inst{1}}

\institute{
  \inst{1} Physics Department, Razi University, Kermanshah
67149, Iran} \pacs{13.60.Hb}{First paces description}
\pacs{12.38.Bx}{Second paces description}\pacs{11.10.Gh}{Three
paces description}

\abstract{We check the impact of the gluon distribution due to the
number of active flavor employed in the calculation of heavy-quark
deep-inelastic scattering (DIS) electro-production on the reduced
cross section and structure functions determined in the leading
order (LO) and next-to-leading order (NLO) quantum chromodynamics
(QCD) analysis using the Hadron Electron Ring Accelerator (HERA)
combined data. The charm and beauty structure functions are
determined and compared with the HERA combined data. The results
are in good agreement with respect to the experimental data. We
also compare the obtained charm and bottom structure functions
with the results from CTEQ6.6 [Phys.Rev.D{\bf78}, 013004(2008)]
and NNPDF3.1[Eur.Phys.J.C{\bf77}, 663(2017)] parameterization
models. The effect of gluon density on these calculations depends
on the number of active flavor at asymptotical values of the
momentum transfer $Q^{2}$. Also, in the HERA kinematic range, the
ratio of $F_{2}^{c\overline{c}}/F_{2}$ and
$F_{2}^{b\overline{b}}/F_{2}$ are obtained. These results and
comparison with the HERA combined data demonstrate
that the suggested method can be applied in analysis of new colliders.\\
}
\begin{document}

\maketitle

\section{I. INTRODUCTION}
Heavy-quark pair production in deeply inelastic scattering (DIS)
provides the predictions of quantum chromodynamics (QCD). These
predictions can be seen in the combined data collected at HERA
[1]. Also, these predictions will also be visible at future
accelerators [2]. The new accelerators (LHeC [3] and FCC-eh [4])
will guide predictable data toward ultra- high energies. In DIS
processes, the heavy-flavor production is dominantly via the
boson-gluon fusion (BGF) process,
$\gamma^{*}g{\rightarrow}\mathcal{Q}\overline{\mathcal{Q}}$, where
$\mathcal{Q}$ is heavy quark. The reaction under study is
\begin{eqnarray}
e^{-}+P~ {\rightarrow}~ e^{-}+\mathcal{Q}\overline{\mathcal{Q}}+X,
\end{eqnarray}
where $P$ is a proton , $\mathcal{Q}\overline{\mathcal{Q}}$ is a
heavy-quark pair and $X$ is any hadronic state allowed. Indeed the
heavy-quark production is sensitive to the gluon distribution, as
in QCD perturbation theory the cross section for the process
$\gamma^{*}P{\rightarrow} \mathcal{Q}\overline{\mathcal{Q}}X$ can
be written as a convolution of the gluon distribution function
$G(x,\mu^{2})$ and the partonic cross section of the photon-gluon
fusion process $\widehat{\sigma}(\gamma^{*}g{\rightarrow}
\mathcal{Q}\overline{\mathcal{Q}})$ [5]. Indeed
$G(x,\mu^{2})=xg(x,\mu^{2})$ which $g(x,Q^{2})$ is the gluon
density. The gluon momentum in the proton is dependent on the
heavy quarks mass. This dependence on the effect of the gluon
density can be seen in both Fixed Flavor Number (FFN)[6] and
Variable Flavor Number (VFN) [7] schemes. Which DIS nucleon
structure functions can be explain using these schemes. In FFNS
heavy quarks are not considered as active. In this case, for
$Q^{2}{\sim}m^{2}_{\mathcal{Q}}$ the heavy flavors are generated
only by BGF. In the FFNS the heavy-quark structure functions at
low values of $x$ are given by
\begin{eqnarray}
F_{2,L}^{\mathcal{Q}\overline{\mathcal{Q}}}(x,Q^{2})=C_{2,L}^{FF,n_{f}}(x,\frac{Q^{2}}{\mu^{2}}){\otimes}
G_{n_{f}}(x,\mu^{2}),
\end{eqnarray}
where $n_{f}=3$ is the number of light quark flavors when all
heavy flavors are considered as massive, and $G_{n_{f}}$ is the
gluon distribution function due to the number of active quark
flavors. Here $C_{2,L}$ are the Wilson coefficients. $\mu^{2}$
denotes the factorization scale and the Mellin convolution is give
by the integral $[f\otimes g](x)=\int_{x}^{1}(dy/y)f(y)g(x/y)$.
For $Q^{2}>m^{2}_{\mathcal{Q}}$ VFNS has been introduced, which
heavy quarks considered as the light-flavor ones. For realistic
kinematics it has to be extended to the case of a General- Mass
VFNS (GM-VFNS) which is defined similarly to the  Zero-Mass VFNS
(ZM-VFNS) in the $Q^{2}/m^{2}_{\mathcal{Q}}{\rightarrow}\infty$
limit[8]. In this scheme, the transition from $n_{f}$ active
flavors to $n_{f}+1$ considered in the construction of charm-quark
parton distribution function (PDF). Rather at some large scale the
transition with two massive quarks (i.e.,
$n_{f}{\rightarrow}n_{f}+2$) has been discussed in Ref.[8]. One of
the ingredients used in the GM-VFNS in the argument of the
respective massless Wilson coefficient functions is rescaling of
the parton momentum fractions $z$ by replacing
$z{\rightarrow}\chi=z(1+\frac{4m_{\mathcal{Q}}^{2}}{Q^{2}})$.
Within these schemes,  the heavy quark densities arise via the
gluon evolution. Gluon behavior in determining the heavy-quark
density depends on the effect of the number of active flavor. In
Ref.[9] this behavior is well illustrated. Authors in [9] showed
an analytical behavior of the gluon distribution function from the
proton structure function due to effects of heavy quarks. Indeed,
accurate knowledge of heavy structure functions due to the effect
of  gluon distribution at small $x$ will play a vital role in
estimating our ability to search for new physics at the LHeC and
FCC-eh colliders.\\
The plane of the present paper is as follows. In section II we
review the heavy flavors influence in gluon density determination,
and we present the gluon distribution functions used in the
present study. In section III we describe the various schemes with
respect to the gluon behavior. The section IV is devoted to the
results and
 discussion. Finally, we give our conclusions in Sec.V.\\

%%%%%%%%%%%%%%%%%%%%%%%%%%%%%%%%%%%%%%%%%%%%%%%%%%%%%%%%%%%%%%%%%
\section{II. Theory}
In this section we briefly present the theoretical   analysis of
the gluon distribution function due to the effect of heavy quarks.
The reader can be refereed to the Ref.[9] for more details. The
DGLAP evolution equations [10] must be modified to take into
account the effect of production thresholds for pairs of charm and
beauty quarks. The proton structure function is defined into the
quark distribution functions $q_{i}(x,Q^{2})$ as
$i{\in}~u,\overline{u},d,\overline{d},s,\overline{s},c,\overline{c},b,\overline{b}$.
The evolution equation for the proton structure function
\begin{eqnarray}
\frac{1}{x}\frac{\partial{F_{2}(x,Q^{2})}}{\partial{{\ln}Q^{2}}}=\frac{\alpha_{s}}{4\pi}
[\int_{x}^{1}\frac{dz}{z^{2}}F_{2}(z,Q^{2})K_{qq}(\frac{x}{z})\nonumber\\
+\sum_{i}e_{i}^{2}\frac{1}{\eta_{i}}\int_{x}^{1}\frac{dz}{z^{2}}
G(\eta_{i}z,Q^{2})K_{gq}(\frac{x}{z})],
\end{eqnarray}
where $e_{i}^{2}$ is the squares of the quark charges, $K^{,}$s
are the splitting functions  and
$\eta_{i}(Q^{2})=1+4\frac{M_{i}^{2}}{Q^{2}}$. For massless quarks
the value of $\eta_{i}$  defined to be 1. In Eq.(3) the gluon
distribution function shift from $z$ to $\eta_{i}z$ for activation
of charm and beauty quarks. The gluon distribution has been
defined by
\begin{eqnarray}
G(x,Q^{2})+\frac{2}{3}\frac{1}{\eta_{c}}G(x_{c},Q^{2})+~~~~~~~~~~~~~~~~~~~~~`
\nonumber\\
~~~~~~~~~\frac{1}{6}\frac{1}{\eta_{b}}G(x_{b},Q^{2})=\mathcal{H}_{3}(x,Q^{2}),
\end{eqnarray}
where
\begin{eqnarray}
\mathcal{H}_{3}(x,Q^{2})=\frac{1}{\omega}\int_{x}^{1}\frac{dz}{z}(\frac{z}{x})^{k}
\sin(\omega{\ln}\frac{z}{x})\mathcal{G}_{3}(x,Q^{2}),
\end{eqnarray}
and
\begin{eqnarray}
\mathcal{G}_{3}(x,Q^{2})=-\frac{3}{4}\frac{4\pi}{\alpha_{s}(Q^{2})}x^{4}
\frac{\partial^{3}({F}_{2}(x,Q^{2})/x)}{{\partial}x^{3}}.
\end{eqnarray}
Here $x_{c}=x(1+4\frac{M_{c}^{2}}{Q^{2}})$ and
$x_{b}=x(1+4\frac{M_{b}^{2}}{Q^{2}})$. In Eq.(5), $k$ and $\omega$
are the real and  imaginary parts of the roots of the factored
form of the differential operator, with $k=-3/2$ and
$\omega=\sqrt{7}/2$. By considering the Laplace transform method,
$\mathcal{H}_{3}(x,Q^{2})$ has been defined by
\begin{eqnarray}
\mathcal{H}_{3}(x,Q^{2})&=&\frac{3}{4}\frac{4\pi}{\alpha_{s}}\{
3{F}_{2}(x,Q^{2})-x\frac{\partial}{\partial{x}}{F}_{2}(x,Q^{2})\nonumber\\
&&-\int_{x}^{1}\frac{dz}{z}{F}_{2}(z,Q^{2})(\frac{x}{z})^{3/2}
[\frac{3}{\omega}\sin(\omega{\ln}\frac{z}{x})\nonumber\\
&&+2\cos(\omega{\ln}\frac{z}{x})]\}.
\end{eqnarray}
where shows that high derivatives of the structure functions do
not appear in $\mathcal{\widehat{H}}_{3}(\nu,Q^{2})$. Here
$\mathcal{H}_{3}(x,Q^{2})$ converted to the $\nu$ form
$\mathcal{\widehat{H}}_{3}(\nu,Q^{2})$ by the substitution
$x=e^{-\nu}$, $\nu=\ln(1/x)$.\\
In the case of four massless quarks, the gluon distribution is
given by the following form
\begin{eqnarray}
G_{4}(x,Q^{2})=\frac{3}{5}\mathcal{H}_{3}(x,Q^{2})=\mathcal{H}_{4}(x,Q^{2})
\end{eqnarray}
When taking mass effects into account, the exact solution for the
gluon distribution at $n_{f}=5$ has the following form
\begin{eqnarray}
G(x,Q^{2})&=&\mathcal{H}_{3}(x,Q^{2})+\sum_{n=1}^{N}(-1)^{n}\sum_{k=0}^{n}\left(
{\begin{array}{c}
n  \\
   k  \\
  \end{array} } \right)\nonumber\\
&&\alpha^{n-k}\beta^{k}\mathcal{H}_{3}(\eta_{c}^{n-k}\eta_{b}^{k}x,Q^{2}).
\end{eqnarray}
Here $\eta_{c}=1+4\frac{M_{c}^{2}}{Q^{2}}$,
$\eta_{b}=1+4\frac{M_{b}^{2}}{Q^{2}}$, $\alpha=(2/3\eta_{c})$,
$\beta=(1/6\eta_{b})$ and $\big( ^{n} _{k} \big)$ is a binomial
coefficient. The summations in Eq.(9) are finite as the sum on $n$
terminates at $N$ such that $(N+1)\ln{\eta_{c}}{\geq}\ln(1/x)$.
Indeed $G_{5}$ was found in the form
\begin{eqnarray}
G_{5}(x,Q^{2})=\frac{6}{11}\mathcal{H}_{3}(x,Q^{2})=\mathcal{H}_{5}(x,Q^{2})
\end{eqnarray}
The fractional uncertainty in $\mathcal{H}_{3}(x,Q^{2})$ was
determined by the fractional statistical error in
$F_{2}(x,Q^{2})$, where $F_{2}(x,Q^{2})$ determined [11] from the
ZEUS structure function data [12]. This fractional statistical
error is defined as
\begin{eqnarray}
\Delta{\mathcal{H}_{3}(x,Q^{2})}=\mathcal{H}_{3}(x,Q^{2})\frac{\Delta{F_{2}(x,Q^{2})}}{F_{2}(x,Q^{2})},
\end{eqnarray}
where the fractional statistical error of the proton structure
function  obtained from the fit parameters shown in Ref.[9].\\

%%%%%%%%%%%%%%%%%%%%%%%%%%%%%%%%%%%%%%%%%%%%%%%%%%%%%%%%%%%%%%%%%
\section{III. Method}
The deeply inelastic heavy-quark structure functions
($F_{k}(x,Q^{2},m_{\mathcal{Q}}^{2})$ for $k=2,L$) are given by
[13] the following forms as
\begin{eqnarray}
F_{k}^{\mathcal{Q}\overline{\mathcal{Q}}}(x,Q^{2},m^{2}_{\mathcal{Q}})=\frac{Q^{2}\alpha_{s}(\mu^{2})}{4m^{2}_{\mathcal{Q}}
\pi^{2}}\int_{x}^{z_{max}}\frac{dz}{z}[e^{2}_{\mathcal{Q}}~g(\frac{x}{z},\mu^{2})c_{k,g}^{(0)}]\nonumber\\
+\frac{Q^{2}\alpha_{s}^{2}(\mu^{2})}{m^{2}_{\mathcal{Q}}\pi}\int_{x}^{z_{max}}\frac{dz}{z}\{e^{2}_{\mathcal{Q}}~g(\frac{x}{z},\mu^{2})(c_{k,g}^{(1)}
+\overline{c}_{k,g}^{(1)}{\ln}\frac{\mu^{2}}{m^{2}_{\mathcal{Q}}})\nonumber\\
+\sum_{i=q,\overline{q}}[e^{2}_{\mathcal{Q}}q_{i}(\frac{x}{z},\mu^{2})(c_{k,i}^{(1)}+\overline{c}_{k,i}^{(1)}{\ln}\frac{\mu^{2}}{m^{2}_{\mathcal{Q}}})\nonumber\\
+e^{2}_{\mathrm{Light},i}q_{i}(\frac{x}{z},\mu^{2})d_{k,i}^{(1)}]\}.\nonumber
\end{eqnarray}
The functions $q_{i}$ for quark and anti-quark denote the parton
densities in the proton. The coefficient functions, defined by
$c_{k,i}^{(l)}$, $\overline{c}_{k,i}^{(l)}$,
($i=g,q,\overline{q};~ l=0,1$) and by $d_{k,i}^{(l)}$,
$\overline{d}_{k,i}^{(l)}$, ($i=q,\overline{q};~ l=0,1$) are
represented in the $\overline{\mathrm{MS}}$-scheme and calculated
in [13]. The gluon density (i.e., $g(x,\mu^{2}$)) is dominant at
low values of $x$, therefore the simplest form after some
manipulation with the expression for
$F_{k}^{\mathcal{Q}\overline{\mathcal{Q}}}(x,Q^{2},m^{2}_{\mathcal{Q}})$
can be written as
\begin{eqnarray}
F_{k}^{\mathcal{Q}\overline{\mathcal{Q}}}(x,Q^{2},m^{2}_{\mathcal{Q}})=2xe_{c}^{2}\frac{\alpha_{s}(\mu^{2})}{2\pi}\nonumber\\
\times\int_{\chi}^{1}\frac{dz}{z^{2}}C_{g,k}
(\frac{x}{z},\frac{Q^{2}}{\mu^{2}})G(z,\mu^{2}),
\end{eqnarray}
where $\chi=\eta_{\mathcal{Q}}x$ and $\eta_{\mathcal{Q}}(Q^{2})=
1+4\frac{m_{\mathcal{Q}}^{2}}{Q^{2}}$. The renormalization scale
is assumed to be $\mu^{2}=4m_{\mathcal{Q}}^{2}+Q^{2}$. In Eq.(12),
$C_{g,k}(\frac{Q^{2}}{\mu^{2}})$ are the scale dependent Wilson
coefficients in the $\overline{\mathrm{MS}}$-scheme with
$C_{g,k}(\frac{Q^{2}}{\mu^{2}})=c_{g,k}$ for $Q^{2}=\mu^{2}$. Here
$C^{\mathcal{Q}}_{g,k}$ are heavy-flavor coefficient functions to
next-to-leading order (NLO) approximation [6,8,13,14] and are
represented in the $\mathrm{\overline{MS}}$ scheme and presented
in the following form
\begin{eqnarray}
C_{k,g}{\rightarrow}C^{0}_{k,g}+\frac{\alpha_{s}(\mu^{2})}{4\pi}[C_{k,g}^{1}
+\overline{C}_{k,g}^{1}\ln\frac{\mu^{2}}{m_{\mathcal{Q}}^{2}}],~~~k=2,L\nonumber
\end{eqnarray}
In the FFNS at low $Q^{2}$, the heavy-flavor structure functions
are given by
\begin{eqnarray}
F_{k}^{\mathcal{Q}\overline{\mathcal{Q}}}(x,Q^{2}\leq
m^{2}_{\mathcal{Q}})=2xe_{c}^{2}\frac{\alpha_{s}(\mu^{2})}{2\pi}\nonumber\\
\times \int_{\chi}^{1}\frac{dz}{z^{2}}C_{g,k}
(\frac{x}{z},\frac{Q^{2}}{\mu^{2}})\mathcal{H}_{3}(z,\mu^{2}).
\end{eqnarray}
In the GM-VFNS at high $Q^{2}$, the heavy-flavor structure
functions are dependence to the active flavor number as
\begin{eqnarray}
F_{k}^{\mathcal{Q}\overline{\mathcal{Q}}}(x,Q^{2}\gg
m^{2}_{\mathcal{Q}})=2xe_{c}^{2}\frac{\alpha_{s}(\mu^{2})}{2\pi}\nonumber\\
\times \int_{\chi}^{1}\frac{dz}{z^{2}}C_{g,k}
(\frac{x}{z},\frac{Q^{2}}{\mu^{2}})G_{n_{f}}(z,\mu^{2});
\end{eqnarray}
where we take $n_{f}=4$ for $m_{c}^{2}<\mu^{2}<m_{b}^{2}$ and
$n_{f}=5$ for $m_{b}^{2}<\mu^{2}<m_{t}^{2}$.\\
In these approaches the observation of heavy-quark pair production
represents a sensitive probe of the gluon density in the proton.
The calculation of
$F_{k}^{\mathcal{Q}\overline{\mathcal{Q}}}(x,Q^{2})$ can be used
to test different gluon distributions. The charm and beauty
structure functions (i.e., $F_{2}^{c\overline{c}}$ and
$F_{2}^{b\overline{b}}$) are also compared to the prediction from
the ZEUS $F_{2}(x,Q^{2})$ experimental data [12]. The global
parameterization of the ZEUS data for the proton structure
function made by authors in [12] and updated by using the data as
combined by the ZEUS and H1 [15] groups in [16]. The explicit
expression for the $F_{2}$ parameterization in a wide range of the
kinematical variables $x$ and $Q^{2}$ obtained   by the following
general form
\begin{eqnarray}
F_{ 2}(x,Q^{2})= D(Q^{2})(1-
x)^{n}\sum_{m=0}^{2}A_{m}(Q^{2})L^{m},
\end{eqnarray}
where the parameters  with their statistical errors are given in
Refs.[11,16]. We can investigate the ratio
$F_{2}^{\mathcal{Q}\overline{\mathcal{Q}}}(x,Q^{2})/F_{
2}(x,Q^{2})$ which this behavior is very interesting in some
evidence from the  EMC experiments [17]. It will be very
interesting to study this effect in the range of available HERA
energy and its extension to future energies in LHeC and FCC-eh.\\
The direct effect of the gluon density with respect to the active
flavor number on the heavy-flavor structure  functions can
indirectly express the reduced cross section behavior. The deep
inelastic heavy-quarks structure functions related to the reduced
cross section are given by
\begin{eqnarray}
\sigma^{\mathcal{Q}\overline{\mathcal{Q}}}_r(x,Q^{2})&=&F^{\mathcal{Q}\overline{\mathcal{Q}}}_2
(x,Q^2)-{\frac{y^2}{Y_+}}F^{\mathcal{Q}\overline{\mathcal{Q}}}_L
(x,Q^2),
\end{eqnarray}
where $y=Q^2/sx$ is the inelasticity with $s$ the ep center of
mass energy squared and $Y_+ =1+(1-y)^2$. With respect to the
mass-threshold, the reduced cross section for heavy quarks at low
$Q^{2}$ is given by
\begin{eqnarray}
\sigma^{\mathcal{Q}\overline{\mathcal{Q}}}_r(x,Q^{2})&=&2xe_{c}^{2}\frac{\alpha_{s}(\mu^{2})}{2\pi}
\times \int_{\chi}^{1}\frac{dz}{z^{2}}[C_{g,2}
(\frac{x}{z},\frac{m_{\mathcal{Q}}^{2}}{Q^{2}})\nonumber\\
&&-{\frac{y^2}{Y_+}}C_{g,L}
(\frac{x}{z},\frac{m_{\mathcal{Q}}^{2}}{Q^{2}})]
\mathcal{H}_{3}(z,\mu^{2}),
\end{eqnarray}
and at high $Q^{2}$ values we have
\begin{eqnarray}
\sigma^{\mathcal{Q}\overline{\mathcal{Q}}}_r(x,Q^{2})&=&2xe_{c}^{2}\frac{\alpha_{s}(\mu^{2})}{2\pi}
\times \int_{\chi}^{1}\frac{dz}{z^{2}}[C_{g,2}
(\frac{x}{z},\frac{m_{\mathcal{Q}}^{2}}{Q^{2}})\nonumber\\
&&-{\frac{y^2}{Y_+}}C_{g,L}
(\frac{x}{z},\frac{m_{\mathcal{Q}}^{2}}{Q^{2}})]
G_{n_{f}}(z,\mu^{2}).
\end{eqnarray}

%%%%%%%%%%%%%%%%%%%%%%%%%%%%%%%%%%%%%%%%%%%%%%%%%%%%%%%%%%%%%%
\section{IV. Results and Discussion}
We studied the sensitivity of the heavy-quark structure functions
and reduced cross sections to the values of the number of active
flavor for a wide range of $Q^{2}$ values. We quantified the
impact of varied $n_{f}$ on the reduced cross section for
$c\overline{c}$ and $b\overline{b}$ production at the BGF
processes at low values of $x$. The effect of the gluon influence
on production of heavy quarks in ep collisions at LHeC and FCC-eh
can be investigated according to the available energy ranges in
the future colliders. Based on the available HERA data, we
investigate this effect on charm and beauty behaviors. In the hard
scale of the scattering processes, the treatment of heavy quarks
requires adapting the number of light flavors in QCD to the
kinematics under consideration. Indeed the heavy quark production
at low values of $x$ via the BGF process is sensitive to the gluon
density in the proton. Dependence on the number of flavors $n_{f}$
in the gluon density behavior, causes different schemes to be
used. At low scales (i.e., $Q^{2}<m_{h}^{2}$), $n_{f}=3$  and at
high scales (i.e., $Q^{2}{\gg}m_{c}^{2}, m_{b}^{2}$), $n_{f}=4$
and 5 are used respectively. The HERA combined data [1] are
obtained with respect to the massive FFNS and different
implementations of the VFNS. The running charm and beauty-quark
masses are obtained as $m_{c}=1.29^{+0.077}_{-0.053}~\mathrm{GeV}$
and $m_{b}=4.049^{+0.138}_{-0.118}~\mathrm{GeV}$, where the
uncertainties are obtained through adding the experimental fit,
model and parameterization uncertainties in quadrature. The strong
coupling constant value  is chosen to be
$\alpha_{s}(M_{z}^{2})=0.118$ [9] . This is matched to the
expression for $n_{f}=5$  giving $\alpha_{s}^{LO}=0.175$ and
$\alpha_{s}^{NLO}=0.150$ and also it is matched in turn to the
expression for $n_{f}=4$ giving
$\alpha_{s}^{LO}=0.182$ and $\alpha_{s}^{NLO}=0.160$.\\
The behavior of the gluon distribution functions with respect to
the number of active flavor $n_{f}$ are shown in Fig.1 at $Q^{2}$
values $10$ and $100~\mathrm{GeV}^{2}$. In this figure the gluon
distribution for $n_{f}=3$ massless quarks compared with the
massless distributions for $n_{f}=4$  in the region
$m_{c}^{2}<Q^{2}\leq m_{b}^{2}$ and $n_{f}=5$ for
$Q^{2}>m_{b}^{2}$. Dash lines are the fractional statistical
errors in $G_{n_{f}}(x,Q^{2})$ due to the Eq.(11), where we used
the fractional statistical errors of proton structure function
from the fit parameters shown in Ref.[11]. The results for
$F_{2}^{c\overline{c}}(x,Q^{2})$ and
$F_{2}^{b\overline{b}}(x,Q^{2})$ with LO and NLO coefficient
functions are depicted for $Q^{2}$ values $6.5, 12, 25, 30, 80$
and $160~\mathrm{GeV}^{2}$ in Figs.2 and 3 respectively. In Fig.2,
the charm structure functions due to the number of active quark
flavors at LO and NLO approximations are shown. In Fig.3, the
bottom  structure functions at LO and NLO have a similar behavior.
These structure functions are shown as a function of $x$ for
different values of $Q^{2}$ and compared with HERA data [18] as
accompanied with total errors. The calculations with respect to
the gluon influence are consistent with the measurements. These
results are also comparable to the phenomenological results of
recent years in Refs.[19,20]. These results for
$F_{2}^{c\overline{c}}(x,Q^{2})$ and
$F_{2}^{b\overline{b}}(x,Q^{2})$ are compared with the
next-to-leading order analysis of the CTEQ6.6 [21] and
NNPDF3.1[22] parameterization models. For $Q^{2}<m^{2}_{b}$ which
it is  below the $b\overline{b}$-production threshold, the
parameterization models have no data as depicted in Fig.3. These
figures indicate that the obtained results from the present
analysis are in good agreement with those obtained by global QCD
analysis of CTEQ and
NNPDF for the heavy quarks structure functions.\\
In Fig.4 we shown the ratio $F_{2}^{c\overline{c}}/F_{2}$ and
$F_{2}^{b\overline{b}}/F_{2}$ where the $F_{2}(x,Q^{2})$
parameterization is taken from Refs.[9,11]. Within the accuracy of
the charm data this ratio seems to be comparable to the
experimental data. The results presented in the measured range for
the three $Q^{2}$  are comparable to the H1 data [23]. We observe
that these ratios are constant at low values of $x$ and increase
with $Q^{2}$ for fixed $x$. We see that this value for the ratio
$F_{2}^{c\overline{c}}/F_{2}$ is approximately between 0.2 and 0.5
in a region of $Q^{2}$ and this prediction is close to the average
result
\begin{eqnarray}
<F_{2}^{c\overline{c}}/F_{2}>=0.237{\pm}0.021^{+0.043}_{-0.039},\nonumber
\end{eqnarray}
in ref.[23]. The ratio $F_{2}^{b\overline{b}}/F_{2}$ is also shown
in this figure (i.e., Fig.4) as it is approximately between 0.009
and 0.03 in a region of $Q^{2}$ values. These ratios, especially
the ratio $F_{2}^{b\overline{b}}/F_{2}$, will be able to be
considered in the LHeC and FCC-eh collisions. In the following,
the error bands illustrated in Figs. 4-7 are due to the heavy
quark mass uncertainties (as it is connected with the choice of
factorization and renormalization scales) and the statistical
errors in the
parameterization of $F_{2}(x,Q^{2})$.\\
In Figs.5 and 6, phenomenological predictions of the charm and
beauty reduced cross sections are compared to the combined HERA
data [1] in a wide range of $Q^{2}$ values.  The center-of-mass
energy ($\sqrt{s}=$318~GeV) for charm and beauty-quark production
used in the combined HERA data. In these figures the  reduced
cross sections are determined with respect to the gluon
parameterization due to the number of active flavor. Our results
compared to the data of the HERA combined [1] at $Q^{2}=2.5, 12,
120$ and $650~\mathrm{GeV}^{2}$. Because the beauty reduced cross
section experimental error in $Q^{2}=2.5~\mathrm{GeV}^{2}$ is very
large, we did not examine this value in Fig.6. The uncertainty of
the reduced cross sections are due to the $F_{2}$ parameterization
as defined in Eq.(11) according to Refs.[11,16]. Consistency
between the determined results with respect to the experimental
data are excellent. \\
The HERA data [1] on heavy-flavor DIS production used of the
ABMP16 set [24]. The ABMP16 PDF fit is based on the FFN and VFN
schemes with respect to the number of active flavor.  Here we use
the Bjorken scaling variable $x$ by replacing $\chi{\rightarrow}x$
in $Q^{2}=sy\chi$.
 In Fig.7, we compared the results of charm and beauty
reduced cross sections in the 4 and 5-flavor gluon distributions
at LO and NLO approximations with HERA combined data [1]. Combined
charm and reduced cross sections (i.e., Figs.6 and 7 in Ref.[1])
compared to the NLO QCD FFNS predictions based on the HERAPDF2.0
FF3A [25] and ABMP16 [24] PDF sets. As shown in Fig.7, the charm
and beauty reduced cross sections of the current analysis at
$Q^{2}$ values $32, 60$ and $200~\mathrm{GeV}^{2}$ are in good
agreement with the HERA combined data. We conclude that influence
of the gluon distribution due to the number of active flavor is
effective in determining heavy quark structure functions and it
will give us hope for the impact of this effect on future
available energies in LHeC and FCC-eh
colliders.\\

%%%%%%%%%%%%%%%%%%%%%%%%%%%%%%%%%%%%%%%%%%%%%%%%%%%%%%%%%%%%%%%%%%%%%%%%%%%%%%%%%%%
\section{V. Summary and Conclusion}
In this paper, we present an application of the method of
extraction of the gluon distribution function due to the number of
active flavor suggested in Ref.[9]. We explored the effect of the
number of active flavor on the reduced cross section of the
heavy-quark production at the BGF processes. The heavy quark
structure functions have been extracted with respect to the gluon
distribution functions within the leading order and
next-to-leading order approximation on the HERA kinematical
region. The obtained explicit expressions for the heavy quark
reduced cross sections are entirely determined by the
$n_{f}$-flavor gluon distribution at asymptotical values of the
momentum transfer $Q^{2}$. Indeed the treatment of heavy quark
structure functions requires adapting the number of light flavors
in QCD to the HERA kinematics under consideration. It has been
found that, at low and high-$Q^{2}$ values in comparison with the
heavy quark mass $m_{\mathcal{Q}}$, NLO results reproduce fairly
well the experimental data. These results show that different
numerical methods are obtained by adopting the $n_{f}$-flavor
gluon distribution scheme variants. Explicitly, we used 4-flavor
gluon distributions for the charm and also 5-flavor gluon
distributions for the bottom and obtained the reduced cross
sections at high $Q^{2}$ values and  at low values of $x$. Also we
have studied bounds on the ratios of
$F_{2}^{\mathcal{Q}\overline{\mathcal{Q}}}/F_{2}$ that are useful
in examining the contribution of heavy quarks
 in the proton in future colliders.\\
%%%%%%%%%%%%%%%%%%%%%%%%%%%%%%%%%%%%%%%%%%%%%%%%

\subsection{ACKNOWLEDGMENTS}

Authors are grateful the Razi University for financial support of
this project. G.R.Boroun would like to thanks H.Khanpour for help
with preparation of the QCD parameterization
models.\\

%%%%%%%%%%%%%%%%%%%%%%%%%%%%%%%%%%%%%%%%%%%%%%%%%%%%%%%%%%%%

%%%%%%%%%%%%%%%%%%%%%%%%%%%%%%%%%%%%%%%%%%%%%%%%%%%%%%%%%%
%%%%%%%%%%%%%%%%%%%%%%%%%%%%%%%%%%%%%%%%%%%%%%%%%%%%%%%%%%
\begin{figure}
\includegraphics[width=.5\textwidth]{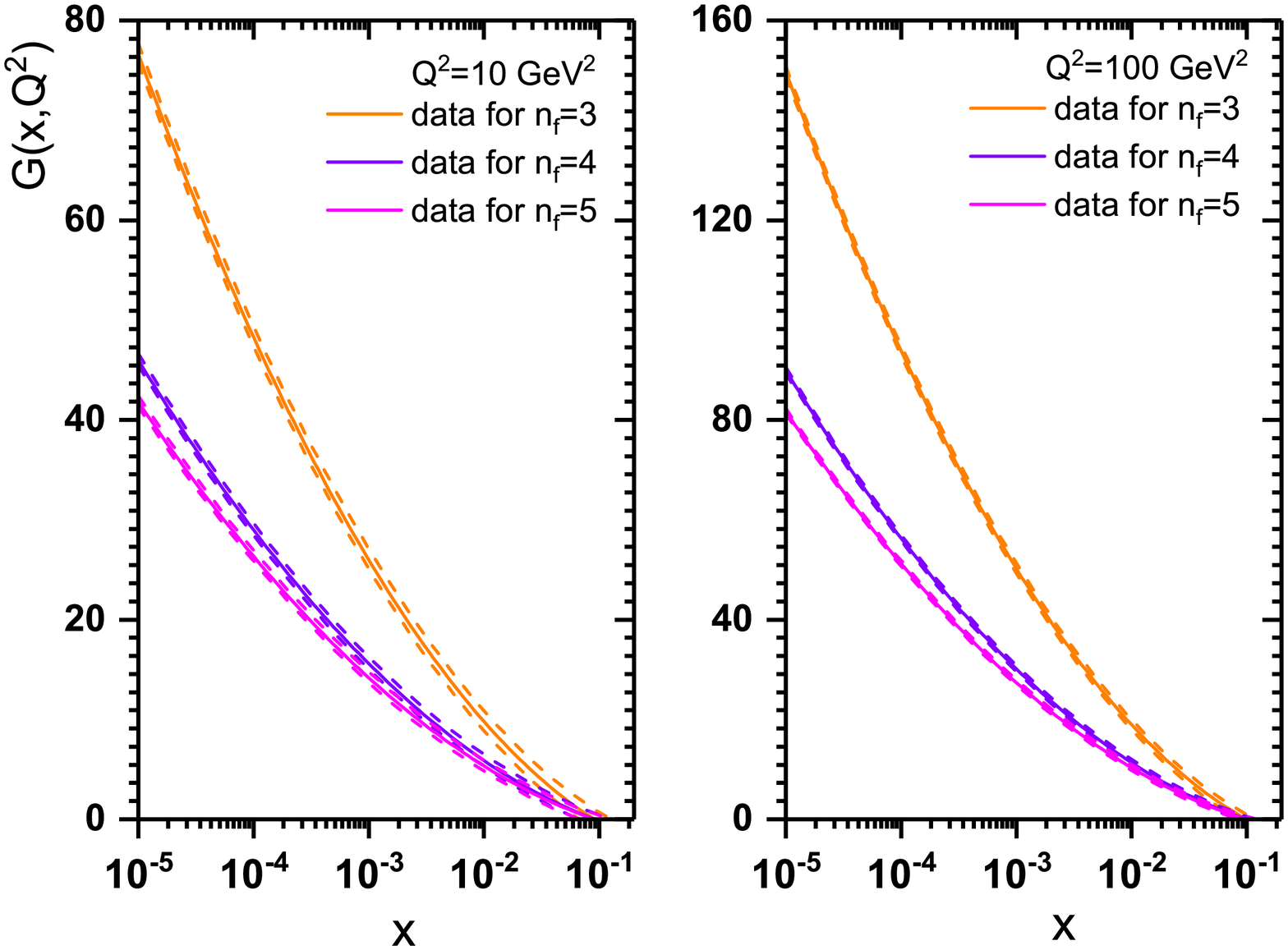}
\caption{A plot of the gluon distribution function vs. x, for two
virtualities due to the number of active flavor. The dash lines
are accompanied with statistical errors. }\label{Fig1}
\end{figure}
\begin{figure}
\includegraphics[width=.55\textwidth]{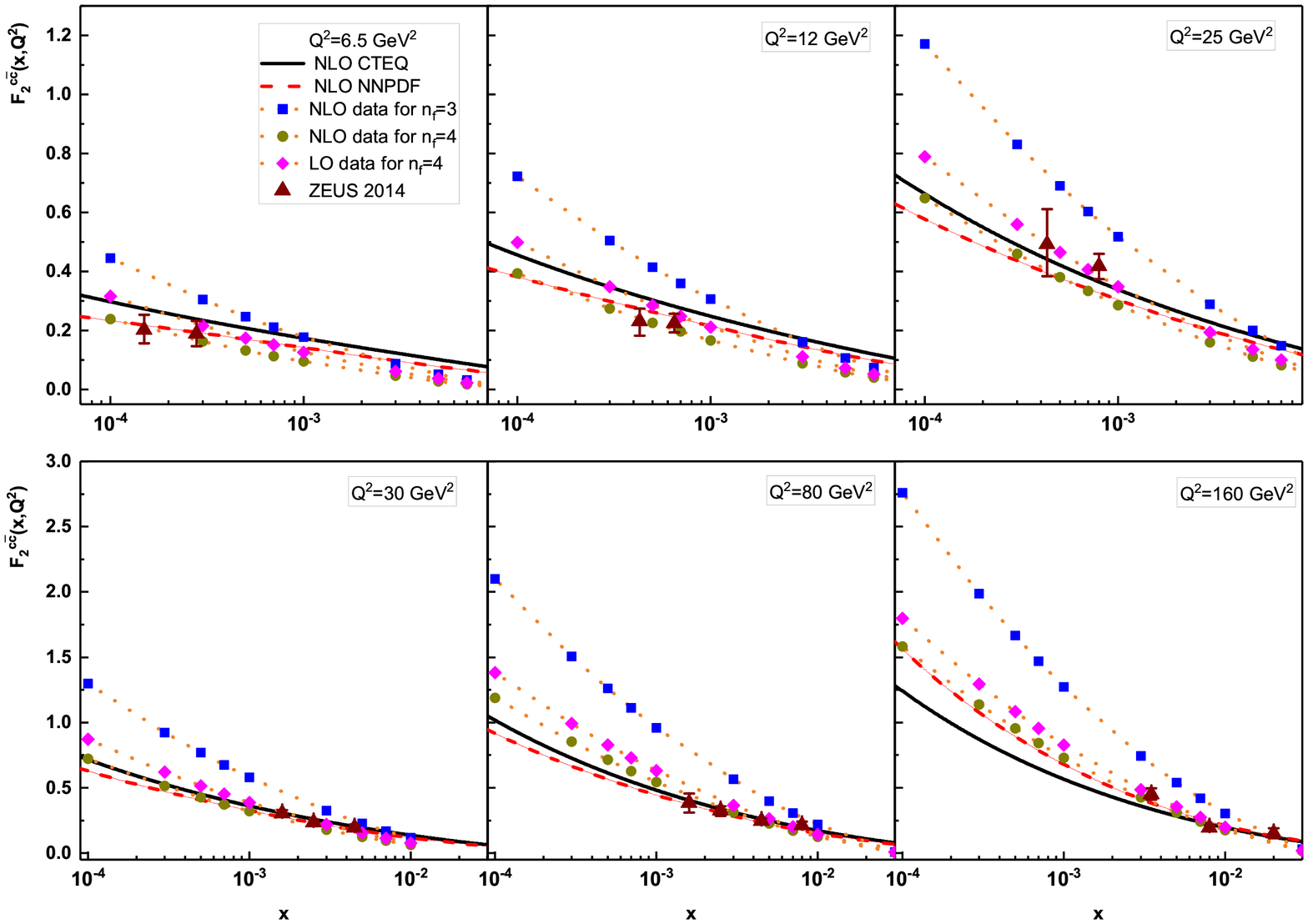}
\caption{The structure function $F_{2}^{c\overline{c}}$ at LO and
NLO approximations  as a function of $x$ for $Q^{2}$ values $6.5,
12, 25, 30, 80$ and $160~\mathrm{GeV}^{2}$. These results are
compared with ZEUS data [18] according to $n_{f}=3$ and $n_{f}=4$.
The experimental data accompanied with total errors. The NLO
results of CTEQ [21] and NNPDF[22] models are also presented
(solid and dashed lines). }\label{Fig1}
\end{figure}
\begin{figure}
\includegraphics[width=.55\textwidth]{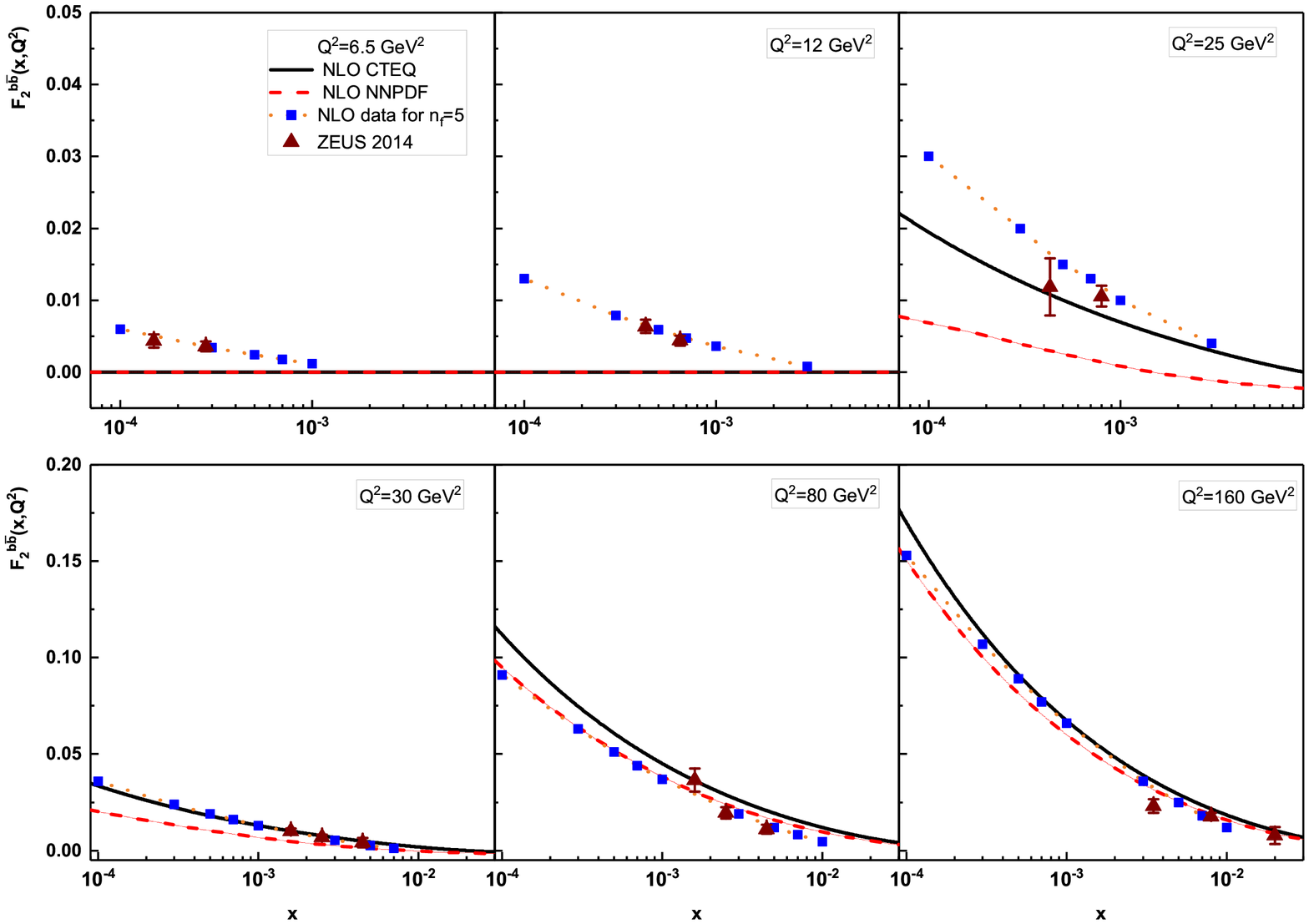}
\caption{The structure function $F_{2}^{b\overline{b}}$ as a
function of $x$ for $Q^{2}$ values $6.5, 12, 25, 30, 80$ and
$160~\mathrm{GeV}^{2}$. These results are compared with ZEUS data
[18] according to $n_{f}=5$. The experimental data accompanied
with total errors. The NLO results of CTEQ [21] and NNPDF[22]
models are also presented (solid and dashed lines)}\label{Fig1}
\end{figure}
\begin{figure}
\includegraphics[width=0.55\textwidth]{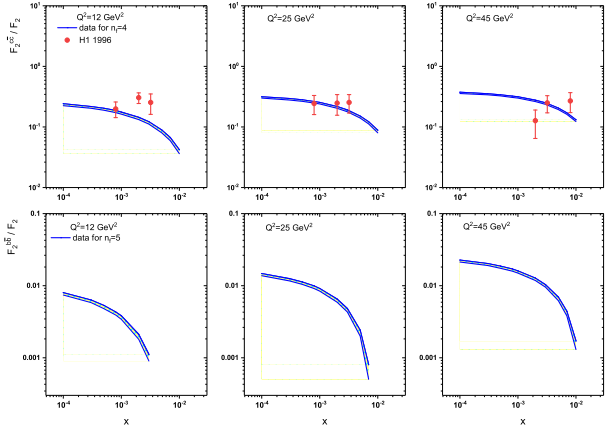}
\caption{The value of the ratio $F_{2}^{c\overline{c}}/F_{2}$ and
$F_{2}^{b\overline{b}}/F_{2}$. Experimental data are taken from
the H1 Collaboration Ref.[23] as accompanied with total error in
quadrature only for the available value of the ratio
$F_{2}^{c\overline{c}}/F_{2}$. }\label{Fig1}
\end{figure}
\begin{figure}
\includegraphics[width=.55\textwidth]{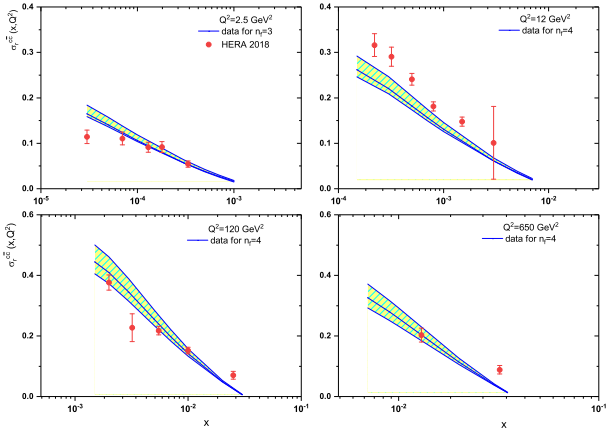}
\caption{Theoretical predictions for $\sigma^{c\overline{c}}_{r}$
 as a function of $x$ at $Q^{2}=2.5, 12, 120$ and $650~\mathrm{GeV}^{2}$ using the
 gluon
parameterization with respect to the number of active flavor.
These curves are associated with  statistical errors due to the
$F_{2}$ parameterization. HERA combined data [1] accompanied with
total errors.}\label{Fig1}
\end{figure}
\begin{figure}
\includegraphics[width=.555\textwidth]{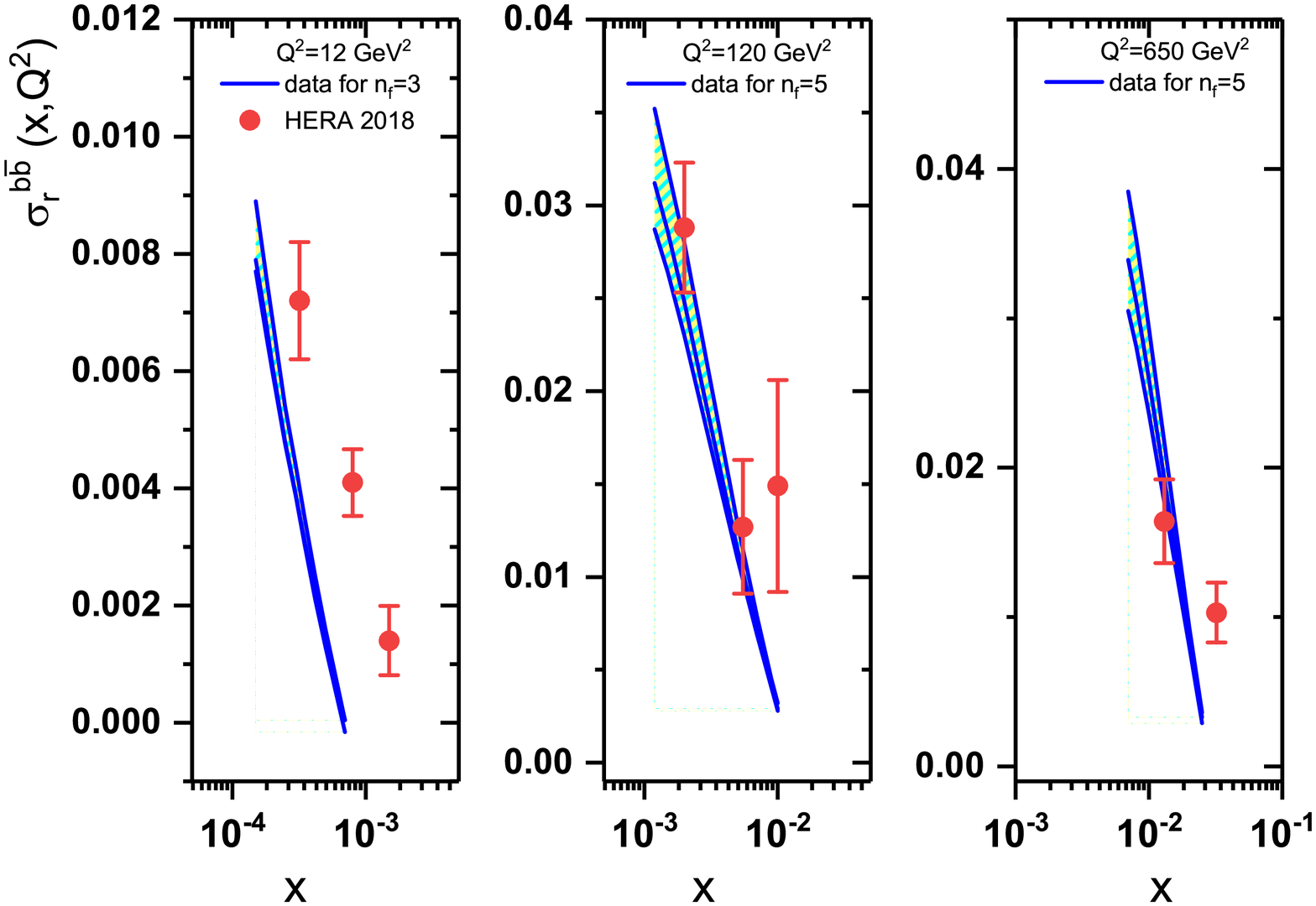}
\caption{The same as Fig.5 for the reduced cross section
$\sigma^{b\overline{b}}_{r}$ .}\label{Fig1}
\end{figure}
\begin{figure}
\includegraphics[width=.55\textwidth]{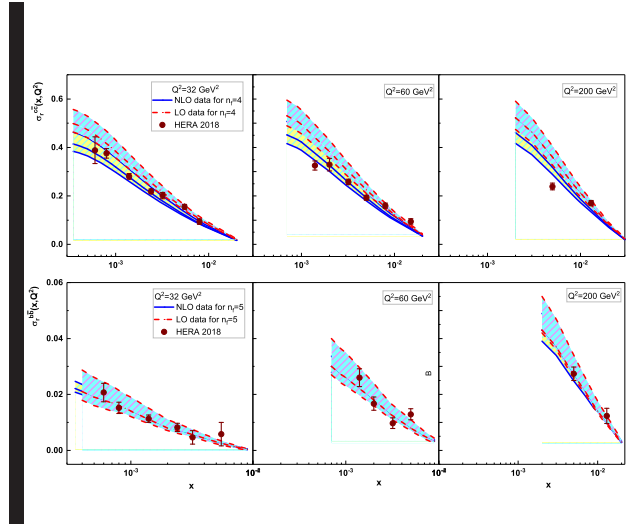}
\caption{The charm and beauty reduced cross sections at LO and NLO
approximations  as a function of Bjorken scaling $x$ for
$Q^{2}=32, 60$ and $200~\mathrm{GeV}^{2}$, compared to the HERA
combined data [1].}\label{Fig1}
\end{figure}

%%%%%%%%%%%%%%%%%%%%%%%%%%%%%%%%%%%%%%%%%%%%%%%%%%%%%%%%%%%%%%%%%%%%%%%%%%%%%%%%%%
%%%%%%%%%%%%%%%%%%%%%%%%%%%%%%%%%%%%%%%%%%%%%%%%%%%%%%%%%

%

\end{document}